\documentclass[]{aastex631}
\usepackage{natbib}
\bibpunct{(}{)}{;}{a}{}{,} 

\usepackage{graphicx}
\usepackage{bm}

\usepackage{amsmath}
\usepackage{tabularx}
\usepackage{multirow}

\usepackage{savesym}
\savesymbol{tablenum}
\usepackage{siunitx}
\restoresymbol{SIX}{tablenum}

\usepackage{placeins}
\usepackage{hyperref}

\DeclareSIUnit[]\rsun
{\text{R\ensuremath{_{\textup{sun}}}}}

\usepackage{booktabs} 

\begin{document} 

    \title{Formation of a Coronal Hole by a quiet-Sun Filament Eruption}
    \author[0000-0001-7662-1960]{Stefan J. Hofmeister}
    \affiliation{Columbia Astrophysics Laboratory, Columbia University, 538 West 120th Street, New York, NY 10027, USA}

    \author[0000-0002-6998-7224]{Eleanna Asvestari}
    \affiliation{Department of Physics, University of Helsinki, P.O. Box 64, 00014, Helsinki, Finland}

    \author[0000-0001-5661-9759]{Karin Dissauer}
    \affiliation{NorthWest Research Associates, 3380 Mitchell Lane, Boulder, CO 80301 USA}
    \affiliation{Institute of Physics, University of Graz, Graz, Universit\"atsplatz 5, 8010 Graz, Austria}
    
    \author[0000-0001-7748-4179]{Michael Hahn}
    \affiliation{Columbia Astrophysics Laboratory, Columbia University, 538 West 120th Street, New York, NY 10027} 

    \author[0000-0002-2655-2108]{Stephan~G.~Heinemann}
    \affiliation{Department of Physics, University of Helsinki, P.O. Box 64, 00014, Helsinki, Finland}
    \affiliation{Institute of Physics, University of Graz, Graz, Universit\"atsplatz 5, 8010 Graz, Austria}

    \author[0000-0003-1862-7904]{Veronika Jercic}
    \affiliation{NASA Goddard Space Flight Center, Greenbelt, MD, USA}

    \author[0000-0002-6232-5527]{Alexandros Koukras}
    \affiliation{Columbia Astrophysics Laboratory, Columbia University, 538 West 120th Street, New York, NY 10027}

    \author[0000-0002-0922-7864]{Kilian Krikova}
    \affiliation{Rosseland Centre for Solar Physics, University of Oslo, PO Box 1029 Blindern, 0315 Oslo, Norway}

    \author[0000-0003-2170-1140]{Jonas Saqri}
    \affiliation{Institute of Physics, University of Graz, Graz, Universit\"atsplatz 5, 8010 Graz, Austria}
    
    \author[0000-0002-1111-6610]{Daniel W. Savin}
    \affiliation{Columbia Astrophysics Laboratory, Columbia University, 538 West 120th Street, New York, NY 10027}

    \author[0000-0003-4867-7558]{Manuela Temmer}
    \affiliation{Institute of Physics, University of Graz, Graz, Universit\"atsplatz 5, 8010 Graz, Austria}

    \author[0000-0003-2073-002X]{Astrid Veronig}
    \affiliation{Institute of Physics, University of Graz, Graz, Universit\"atsplatz 5, 8010 Graz, Austria}
     \affiliation{Kanzelh\"ohe Observatory for Solar and Environmental Research, Kanzelh\"ohe 19, University of Graz, Graz, Austria}

   \date{\today}

  \begin{abstract}
    A coronal hole formed as a result of a quiet-Sun filament eruption close to the solar disk center on 2014~June~25. We studied this formation using images from the Atmospheric Imaging Assembly~(AIA), magnetograms from the Helioseismic and Magnetic Imager~(HMI), and a differential emission measure~(DEM) analysis derived from the AIA images. The coronal hole developed in three stages: (1)~formation, (2)~migration, and (3)~stabilization. In the formation phase, the emission measure (EM) and temperature started to decrease six hours before the filament erupted. Then, the filament erupted and a large coronal dimming formed over the following three hours. Subsequently, in a phase lasting $15.5$~hours, the coronal dimming migrated by \SI{\approx 150}{\arcsec} from its formation site to a location where potential field source surface extrapolations indicate the presence of open magnetic field lines, marking the transition into a coronal hole. During this migration, the coronal hole drifted across quasi-stationary magnetic elements in the photosphere, implying the occurrence of magnetic interchange reconnection at the boundaries of the coronal hole. In the stabilization phase, the magnetic properties and area of the coronal hole became constant. The EM of the coronal hole decreased, which we interpret as a reduction in plasma density due to the onset of plasma outflow into interplanetary space. As the coronal hole rotated towards the solar limb, it merged with a nearby pre-existing coronal hole. At the next solar rotation, the coronal hole was still apparent, indicating a lifetime of $>1$~solar rotation.


   \end{abstract}


\section{Introduction}

Coronal holes are one of the primary structures of the solar corona. They have a lower density and temperature compared to the ambient corona, which makes them appear dark in coronal images at extreme ultraviolet (EUV) and soft X-ray wavelengths  \citep{Landi2008,Hahn2011,saqri2020,heinemann2021}.  A given coronal hole can often be observed for several solar rotations and sometimes grows to cover a significant fraction of the coronal surface \citep{Heinemann2020}. Additionally, coronal holes are characterized by a magnetic flux topology that is open to interplanetary space, along which solar plasma is accelerated to supersonic speeds forming high-speed solar wind streams that traverse interplanetary space \citep[][]{nolte1976, cranmer2009}. 

Although coronal holes are striking features in coronal images, their detailed formation process is rarely observed. Direct observations have only been reported by \citet{solodyna1977} using observations from the S-054 X-ray spectrographic telescope on board Skylab, by \citet{karachik2010} and \citet{yang2010} using observations from the Extreme-ultraviolet Imaging Telescope (EIT) on board the Solar and Heliospheric Observatory (SOHO), and by \citet{inglis2019} using observations from the Atmospheric Imaging Assembly (AIA) on board the Solar Dynamics Observatory (SDO). These studies reported that coronal holes formed within \SI{6}{hours} to \SI{3.4}{days} and that the coronal emission in the EUV and X-ray decreased by $25$~to~\SI{60}{\percent}, as compared to the quiet-Sun values. The coronal holes studied by \citet{karachik2010}, and probably the one by \citet{yang2010}, formed from the decay of active regions, while the coronal holes from \citet{inglis2019} formed within the quiet Sun. The coronal hole from \citet{solodyna1977} possibly formed by a filament eruption, as they noted the disappearance of an H$\alpha$ filament accompanied by the appearance of a nearby coronal dimming prior to the appearance of the coronal hole. 

The lack of further direct observations is due to the unique nature of coronal holes and their observational limitations. Newly formed coronal holes are small; and when they are not close to the solar disk center, they are easily obscured by brighter neighboring structures. Indirect studies based on the statistical locations of coronal holes and based on coronal simulations find that polar coronal holes result from meridional magnetic flux transport towards the poles \citep{bohlin1978, wang2009}, whereas low-latitude coronal holes likely form from decaying active regions \citep{timothy1975, fainshtein2004, petrie2013} and filament eruptions \citep{fainshtein2004}. 

Filament eruptions can also produce coronal dimmings, which may resemble small low-latitude coronal holes in coronal observations but are physically distinct phenomena. These coronal dimmings are localized, transient decreases of emission in EUV and soft X-rays that form in the wake of filament eruptions and coronal mass ejections, due to the expansion and plasma evacuation of the erupting structure \citep[e.g.,][]{Sterling:1997, Thompson:2000}. Dimmings associated with the expansion of the flux rope feet and the evacuation of plasma therein are generally termed core or flux rope dimmings \citep{Sterling:1997, Veronig:2025}. These core dimmings can temporarily reach EUV intensities as low as typically seen in coronal holes \citep[e.g.,][]{Attrill:2005} and have therefore historically been referred to as “transient coronal holes” \citep{Rust:1983, Attrill:2008}. However, unlike true coronal holes, the intensities of coronal dimmings recover gradually to quiet-Sun values. Coronal dimmings persist on average only for \SI{8} to \SI{11}{hours}, with the longest-lived cases observed lasting  less than four days \citep{Attrill:2008, reinard2008, krista2017, krista2022, Ronca_2024}. 


On 2014~June~25, SDO captured in great detail the formation of a coronal hole by a large quiet-Sun filament eruption at the solar disk center: SDO observed the filament eruption; a coronal dimming; and the subsequent evolution of the dimming into a small coronal hole, which then merged with a nearby coronal hole. The newly formed portion of the merged coronal hole was again visible at disk center at the next solar rotation, setting its lifetime to $>1$ solar rotation. The position at the solar disk center during the formation phase minimized any obscuration effects from brighter, nearby coronal structures. Furthermore, the impulsive nature of the filament eruption caused the formation and early evolution of the coronal hole to unfold rapidly and clearly. These features enable us to study the formation of this coronal hole in unprecedented detail. Here, we investigate the formation and early evolution of the coronal hole, analyzing its area, temperature, emission measure, magnetic evolution, and migration away from its birth location to a region characterized by modeled open magnetic fields. The observation of the formation, migration, and stabilization phases spanned one and a half days.

The rest of this paper is structured as follows: Section~\ref{sec:datasets} introduces the datasets and the applied methods. Section~\ref{sec:formation_and_evolution} analyzes the formation and early evolution of the coronal hole. Section~\ref{sec:discussion} discusses the results in terms of underlying physical processes. Section~\ref{sec:summary} summarizes our findings.

\section{Datasets and Methods} \label{sec:datasets}

We studied the solar atmosphere in the AIA~$171$, $193$, $211$,~and \SI{304}{\angstrom} channels, collected at a cadence of \SI{5}{min} \citep{lemen2012, pesnell2012}. All AIA images were rebinned by a factor of $4 \times 4$ to increase the signal-to-noise ratio; corrected for instrumental short-to-long distance scattered light using the revised point-spread functions for AIA  with the basic iterative deconvolution algorithm \citep{hofmeister2024a, hofmeister2025}; and normalized by their exposure time. The coronal observations in the AIA~$171$, $193$,~and \SI{211}{\angstrom} filtergrams primarily show emission from \ion{Fe}{9} at \SI{0.8}{MK}, \ion{Fe}{12} at \SI{1.6}{MK}, and \ion{Fe}{14} at \SI{2}{MK}, respectively. To identify the filament, we used AIA~\SI{304}{\angstrom} filtergrams, which show \ion{He}{2} emission at transition region temperatures of \SI{50000}{K}. 

We identified the boundaries of the coronal hole from June~25 10:30 to June~26~18:00 using a thresholding technique on AIA~\SI{193}{\angstrom} filtergrams, similar to \cite{rotter2012}, \citet{hofmeister2017}, \citet{heinemann2018a}. All times given in this study are in universal time (UT). After June~26~18:00, the location of the coronal hole boundaries could not be robustly identified anymore in the glare of the surrounding quiet-Sun loops. We manually adjusted the threshold for each image to best match the apparent coronal hole boundary in the image. To account for the human bias involved, we determined two sets of boundaries: a compact one comprising the core of the coronal hole and a wider boundary, which  also includes connected darker areas during the migration phase.  When not explicitly otherwise stated, we refer in our analysis to the compact coronal hole boundaries. Furthermore, to establish a baseline for the evolution of coronal plasma properties, we also examine the plasma properties within the coronal hole boundary for the \SI{6}{hours} prior to its formation. We calculated the evolution of the coronal-hole-averaged intensities, its projection-corrected area, and the position of its center of mass, following \citet{hofmeister2017} and \citet{heinemann2018a}.

We derived the plasma temperature $T$, differential emission measure (DEM), and emission measure (EM) of the coronal hole from the AIA filtergrams in the~$171$, $193$, and~\SI{211}{\angstrom} channels using the DEM code of \citet{cheung2016}. The DEM is a measure of the temperature-resolved coronal plasma content and the EM a measure of the temperature-integrated plasma content. We employed the temperature response functions provided by the AIA team for coronal abundances, set Gaussian basis functions for the DEM code, derived the DEM using $100$ linear temperature bins in the temperature range of $0.3$--\SI{3}{MK}, and adjusted the fitting precision so that the simulated AIA counts, derived by multiplying the DEM with the temperature response functions of AIA, matched the observed AIA counts to within \SI{10}{\percent}. We excluded the AIA $94$, $131$, and \SI{335}{\angstrom} channels from the inversions, as their signals were in the noise regime; and their inclusion prevented the DEM from converging to a physically meaningful solution. We neglected any possible changes in coronal elemental abundances during the coronal hole formation, as no spectroscopic measurements of the abundances were available. As a consequence, as commonly done, we also neglected the first-ionization potential (FIP) effect, according to which Fe~abundances are typically decreased in coronal holes as compared to quiet-Sun regions \citep{feldman2002}. 
Because this possible decrease in Fe abundance was neglected, the derived EM for the coronal hole should be regarded as a lower bound. The temperature estimate is largely unaffected, because the Fe abundance cancels out in the temperature determination. 
To study the evolution of the coronal plasma density and temperature during the formation of the coronal hole, we averaged the DEM and EM over its area.

Using the DEM, we tested for adiabatic cooling of the plasma during the filament eruption and formation of the coronal hole. Adiabatic cooling refers to the reduction in plasma temperature resulting from the plasma expansion under the condition that no heat is exchanged with the surrounding environment. This adiabatic expansion might occur in the expanding magnetic fields of the rising, erupting filament. Adiabatic cooling follows the relation
\begin{equation}
    T\, n^{\gamma - 1} = \text{constant}, \label{eq:adiabatic_raw}
\end{equation}
where $n$ is the plasma particle number density, comprising electrons and ions in the corona; and $\gamma = 5/3$ is the adiabatic constant. The plasma particle number density is related to the electron number density by $n  =   k\, n_e$, where $k$ is a constant related to the abundances and ionization states and which will cancel out in the following steps. 
The electron number density $n_e$ is related to the EM by 
\begin{equation}
    EM = \int n_e^2 \, dl, \label{eq:dem}
\end{equation}
where the integral extends along the line-of-sight $l$. 
When we denote the initial time as $t_0$ and a subsequent time as $t_1$ and assume that the plasma can be modelled by a single temperature, it follows from Equations~(\ref{eq:adiabatic_raw}) and~(\ref{eq:dem}) that the evolution of $T$ for plasma that undergoes adiabatic cooling obeys
\begin{equation}
   \frac{T_{t_1}}{T_{t_0}}  = \left(\frac{EM_{t_1}}{EM_{t_0}}\right)^\frac{\gamma - 1}{2}. \label{eq:adiabatic}
\end{equation}
Using this equation and the temperature and EM timelines derived from the DEMs, we evaluated whether the cooling of the plasma during the filament eruption and coronal hole formation was consistent with adiabatic cooling. 

We derived the average photospheric magnetic flux density and the total signed magnetic flux within the coronal hole boundaries using the low-noise photospheric line-of-sight magnetograms taken by the Helioseisemic and Magnetic Imager \citep[HMI; ][]{scherrer2012, schou2012} on board SDO. We also refer to the total magnetic flux of the coronal hole as its ``open'' magnetic flux since the flux extends far into interplanetary space. We assumed the magnetic field in coronal holes to be statistically radial and therefore transformed the line-of-sight magnetic field to a radial magnetic field following \citet{hofmeister2017}. This avoids systematic errors in the magnetic flux measurements due to the Sun's rotation over the period of the study. Furthermore, we identified photospheric magnetic elements as regions with an absolute magnetic field strength greater than \SI{50}{G} in the magnetograms.
    
We accompanied our observational analysis with coronal magnetic field reconstructions at 2014~June~25~08:00 and around July~22~08:00 using the potential field source surface (PFSS) model implemented in the European Heliospheric Forecasting Information Asset \citep[EUHFORIA;][]{pomoell2018}. The model solves the Laplace equation for a potential coronal magnetic field by employing a numerical method of expansion in solid harmonics, expressed in spherical coordinates. To initialize the model, we used global magnetic field maps at a resolution of \SI{1}{\degree}, which were based on HMI observations from the solar rotations around 2014~June~25~08:00 and around July~22~08:00, respectively, processed by the Air Force data assimilative photospheric flux transport model \citep[ADAPT;][]{Arge10, Hickmann2015, barnes2023}, and smoothed by a Gaussian smoothing with a $\sigma=\SI{0.8}{\degree}$. The outer boundary of the PFSS model, known as the source surface where the magnetic field is assumed to be radial, was placed at 1.6 solar radii. To identify open magnetic field regions, we calculated the magnetic field up to the solid harmonics degree $\ell$ = 140 and then traced the magnetic field lines from the source surface to the photosphere and vice versa.  Finally, we projected the open-field maps to the field of view of AIA.

\section{The Formation and Early Evolution} \label{sec:formation_and_evolution}
\FloatBarrier

\begin{figure}
    \centering
    \includegraphics[width=1.\linewidth]{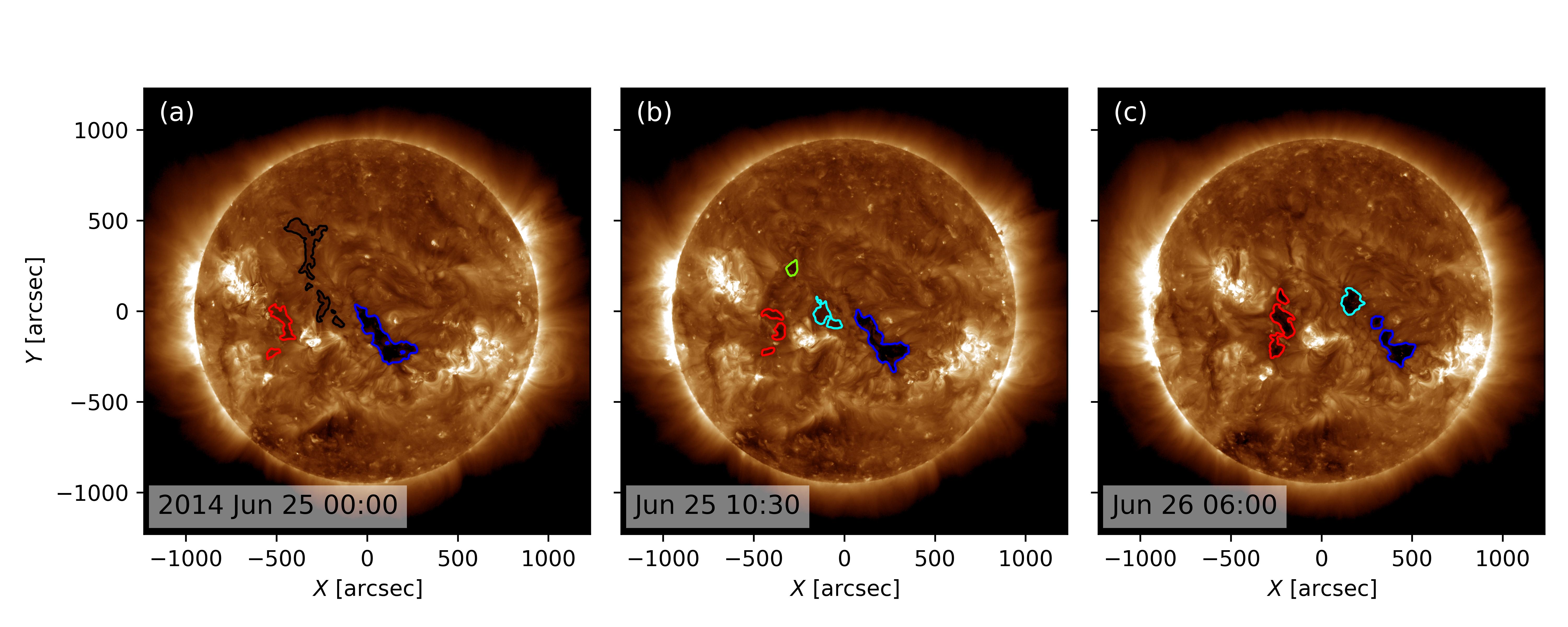}
    \caption{Evolution of the solar corona during the coronal hole formation, as observed in AIA~\SI{193}{\angstrom}. (a)~June~25~00:00: a large quiet-Sun filament channel (marked in black) is present east of the central meridian. A coronal hole (blue) lies to its west, and a dim region (red), possibly another coronal hole obscured by overlying quiet-Sun loops, is visible to the east. (b)~10:30: the filament erupts, producing two coronal dimmings (cyan and green). (c)~June~26~06:00:  The cyan dimming migrates north-westward and develops into the coronal hole under investigation. The green dimming recovers to quiet-Sun intensities.}
    \label{fig:large_overview}
\end{figure}


\begin{figure}
    \centering
    \includegraphics[width=.8\linewidth]{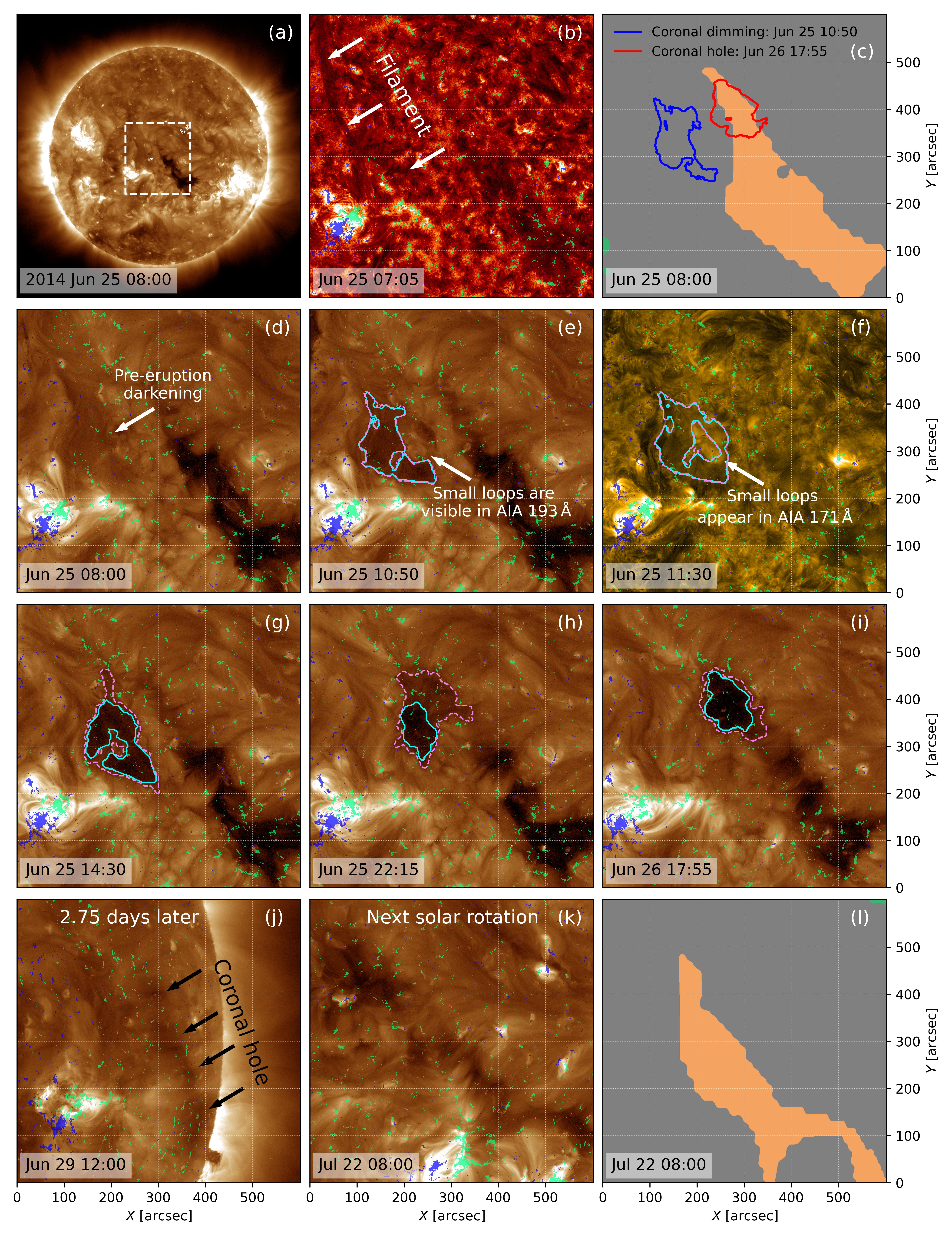}
    \caption{Overview of the coronal hole formation. (a)~Full-disk AIA~\SI{193}{\angstrom} image, with the field of view for subsequent panels outlined in white. In the following panels, the field of view is adjusted to follow the solar rotation over time.
  (b)~Filament channel before eruption as seen in AIA~\SI{304}{\angstrom}. (c)~Open magnetic flux regions from PFSS extrapolations, with orange for positive flux and green for negative flux. The blue and red contours mark coronal hole boundaries at June~25~10:50 and June~26~17:55, respectively. (d)-(e) and (g)-(i)~Coronal dimming and its subsequent evolution into a coronal hole as observed in AIA~\SI{193}{\angstrom}, and (f)~in AIA~\SI{171}{\angstrom}. (i)~The coronal hole $2.75$~days later, close to the solar limb, as observed in AIA~\SI{193}{\angstrom} observations. (k)-(l)~The coronal hole near disk center at the next solar rotation, as observed in AIA~\SI{193}{\angstrom} observations and PFSS extrapolations, respectively. In the EUV observations, photospheric magnetic elements of positive and negative polarity are overlaid in green and blue, respectively. The compact and wider coronal hole boundaries, which are used for the study of the evolution of the coronal hole properties, are marked in cyan and violet, respectively. An associated movie showing the evolution is available online.}

    \label{fig:overview}
\end{figure}

\begin{table}[t]
    \centering
    \begin{tabular}{l|c c c c c c}
                                  & \multirow{2}{*}{Quiet Sun} &  End of      & End of       & End of & End of \\
        Parameter                 &                               & Precursor    & Eruption     & Movement & Stabilization \\ \hline
        $A$~[$10^{9}$~km$^2$] &              $5$ -- $6$                       & $8$ -- $9$     & $7$ -- $10$    & $5$ -- $8$     & $5$ -- $8$     \\
        $B$~[G]               &              $2.5$ -- $2.4$              & $2.3$ -- $2.3$ & $2.5$ -- $2.6$ & $3.5$ -- $3.7$ & $4.3$ -- $4.4$ \\
        $\Phi$~[$10^{20}$~Mx]  &             $1.4$ -- $1.5$                  & $2.0$ -- $2.1$ & $1.9$ -- $2.6$ & $1.7$ -- $3.0$ & $2.0$ -- $3.4$ \\
        AIA~\SI{171}{\angstrom}~[DN] &       $103$ -- $103$              & $117$ -- $119$ & $115$ -- $127$ & $101$ -- $115$ & $76$ -- $97$ \\
        AIA~\SI{193}{\angstrom}~[DN] &       $153$ -- $153$              & $77$ -- $80$ & $53$ -- $66$ & $34$ -- $54$ & $27$ -- $46$ \\
        AIA~\SI{211}{\angstrom}~[DN] &       $65$ -- $65$              & $30$ -- $32$ & $22$ -- $26$ & $13$ -- $18$ & $11$ -- $17$ \\
        EM~[$10^{26}$~cm$^{-5}$]     &       $2.3$ -- $2.3$             & $1.4$ -- $1.5$ & $1.1$ -- $1.3$ & $0.8$ -- $1.1$ & $0.6$ -- $0.9$ \\
        $T$~[MK]                       &        $1.7$ -- $1.7$              & $1.5$ -- $1.5$ & $1.3$ -- $1.4$ & $1.1$ -- $1.2$ & $1.2$ -- $1.3$ \\
    \end{tabular}
    \caption{Coronal hole parameters at different evolutionary stages. The stages here are the quiet-Sun location within the first coronal hole boundary before the coronal hole was formed (2014~June~25~06:00), the end of the precursor phase  (2014~June~25~10:30), the end of the eruption phase  (2014~June~25~13:30), the end of the migration phase  (2014~June~26~05:00), and the end of the stabilization phase  (2014~June~26~18:00). The parameters are area $A$, average magnetic flux density $B$, total magnetic flux $\Phi$, average intensity in the AIA~\SI{171},~193, and \SI{211}{\angstrom}~channels in data numbers~[DN], EM, and temperature $T$. In each cell, the first number gives the parameter derived from the compact boundaries and the second number the parameter derived from the wider boundaries. }
    \label{tab:chparameters}
\end{table}

\begin{figure}
    \centering
    \includegraphics[width=.9\linewidth]{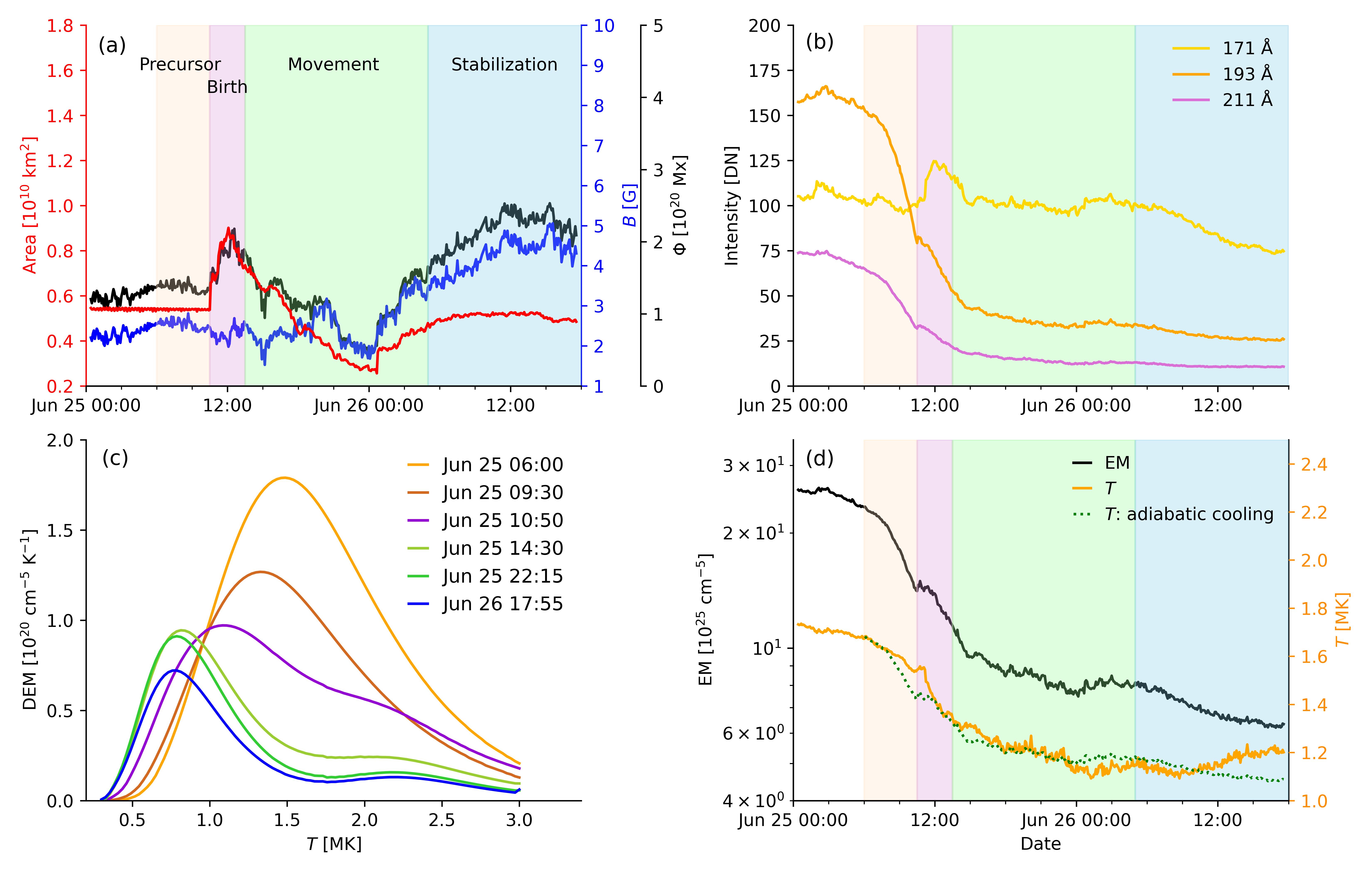}
    \caption{Evolution of the coronal hole properties as derived from the compact coronal hole boundaries. (a)~Area, average magnetic flux density $B$, and total magnetic flux $\Phi$. (b)~Average AIA~$171$,~$193$, and~\SI{211}{\angstrom} intensities. (c)~DEM for selected six times. (d)~EM,~T, and expected temperature from adiabatic cooling.}
    \label{fig:timelines}
      \includegraphics[width=.9\linewidth]{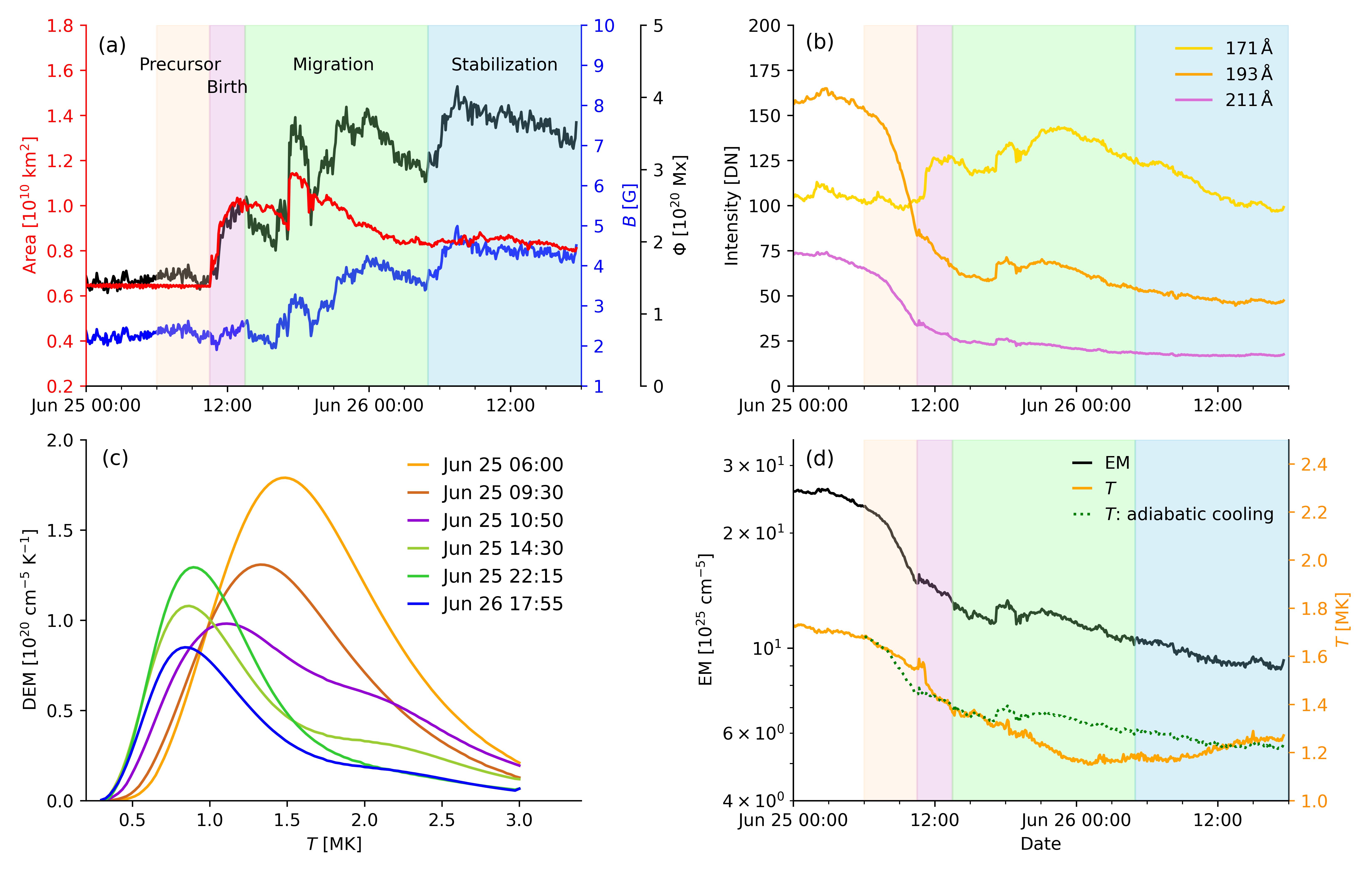}
    \caption{Same as Fig.~\ref{fig:timelines}, but derived from the wider coronal hole boundaries.}
    \label{fig:timelines2}
\end{figure}

On 2014~June~25, a large, north–south oriented quiet-Sun filament channel was observed east of the solar central meridian, as outlined in black in Figure~\ref{fig:large_overview}(a). West of the filament channel, a coronal hole was present as outlined in blue, while to the east lay a dim, ambiguous structure, possibly another coronal hole partially obscured by overlying quiet-Sun loops, as outlined in red. Between~10:00 and~11:00, the filament in the filament channel erupted, producing a coronal mass ejection (CME) with a velocity of \SI{489}{km\,s^{-1}} as reported in the continuously updated SOHO/LASCO CME catalog \citep{gopalswamy2009}. Two coronal dimmings formed at the footpoints of the filament. The western dimming, shown in panel~(b) in cyan and located near disk center, was unobstructed in AIA observations. The dimming migrated north-westward and developed into the coronal hole under investigation. The eastern dimming, shown in green, recovered within \SI{6}{hours} to quiet-Sun intensities.

The formation of the western dimming and its evolution into a coronal hole is detailed in Figure~\ref{fig:overview} and the associated movie. The figure shows snapshots in the AIA~304,~171, and~\SI{193}{\angstrom} channels and the PFSS extrapolation for the magnetic field topology. The AIA images are overlaid with the location of photospheric magnetic elements. In Figure~\ref{fig:timelines} and~\ref{fig:timelines2} and Table~\ref{tab:chparameters}, we show the associated evolution of the coronal hole area~$A$; average magnetic flux density~$B$; total magnetic flux~$\Phi$; average intensity in the AIA~171,~193, and~\SI{211}{\angstrom} channels; EM; and temperature~$T$. 

Figure~\ref{fig:overview}(a) identifies the field of view of the study. Within this field of view, the western leg of the filament is visible in AIA~\SI{304}{\angstrom}, as shown in panel~(b). The footpoint of this leg is about \SI{150}{\arcsec} south-east of a location where the PFSS model suggests the presence of open magnetic field, shown in panel~(c). The pre-existing western coronal hole seen in Figure~\ref{fig:large_overview} is also located within this region of modeled open magnetic field. On June~25~06:00, the filament started to rise and the coronal emission at the footpoint of the filament decreased, which is shown in panel~(d). Between~10:00 and~11:00, the filament erupted and led to the rapid formation of a coronal dimming. In panels~(e)--(i), we show the dimming evolving into the coronal hole under study. For simplicity, we will call it a coronal hole from the time when the dimming has fully developed, i.e., from 13:30 on, since we cannot determine the exact time when the structure is no longer a coronal dimming and has evolved into a coronal hole. This choice will be discussed in more detail in the discussion in Section~\ref{sec:discussion}. From 13:30, the newly formed coronal hole started to migrate towards the north-west. On June 26~05:00, it reached its final location at the northern end of the region with modeled open magnetic field lines. It then stabilized there close to the pre-existing coronal hole and later merged with this pre-existing coronal hole. It can be identified for at least $2.75$~more days as it rotated towards the western solar limb, i.e., at least until $4$~days after its formation, which is shown in panel (j). At the next solar rotation, the coronal hole is still apparent in EUV observations at disk center and in the PFSS extrapolations, which sets its lifetime to $>1$~solar rotation and is shown in panels (k)-(l).

Based on these observations, we divide the formation and early evolution of the coronal hole in three phases: (1)~the formation phase from June~25~06:00 to~13:30, (2)~a subsequent migration phase to the location of the open magnetic field until June~26~05:00, and (3)~a stabilization phase until~18:00. In the following, we analyze these three phases in more detail.

\subsection{Phase 1: Formation} \label{sec:formation}

The formation phase consists of two parts: a precursor phase  from June~25, 06:00~to 10:30, and a  birth phase from~10:30~to~13:30. The filament eruption occurs between these phases from about 10:00~to~11:00. 

In the precursor phase, the filament rose, and even before it had fully erupted, the EUV corona at the footpoint of the filament started to darken. This darkening will later become the coronal dimming and subsequently the coronal hole under investigation. The timelines in Fig.~\ref{fig:timelines} show that this darkening corresponds to AIA~\SI{193}{\angstrom} and~\SI{211}{\angstrom} emission levels decreasing to \SI{51}{\percent} and \SI{49}{\percent}, respectively, as compared to the prior quiet-Sun values, while the AIA~\SI{171}{\angstrom} intensity remained approximately constant. The corresponding EM and temperature timelines indicate that the EM decreased from $2.3$~to~\SI{1.4e26}{cm^{-5}} and the average plasma temperature decreased from $1.7$~to~\SI{1.5}{MK}. This temperature decrease relative to the EM decrease follows the relation expected from adiabatic cooling, given by Equation~(\ref{eq:adiabatic}). Therefore, the darkening of the coronal plasma in this precursor phase might be interpreted as an effect of adiabatic expansion during the rising of the filament. As the filament rises, its magnetic field becomes primarily radially stretched away from the photospheric footpoints. The plasma in the filament, which was originally held down due to magnetic confinement within the stable filament, expands vertically due to the lower ambient magnetic and gas pressure in the higher corona. The expansion causes the plasma to cool down, leading to the decreased temperature and emission.

In the birth phase that followed, the filament fully erupted as seen in the movie in the AIA~\SI{304}{\angstrom} channel. The darkening in AIA~\SI{193}~and \SI{211}{\angstrom}~channels continued, and the coronal dimming formed. Starting from~10:30, the dimming could be clearly distinguished from its surroundings. This is the time that we began to track the evolution of the area of the feature. The timelines in Fig.~\ref{fig:timelines} show that the size of this coronal dimming rapidly increased to an area of \SI{9e9}{km^2}, with a growth rate of about \SI{6e9}{km^2\,h^{-1}} for \SI{1.5}{hours}. The average magnetic flux density in the area encompassed by the coronal dimming remained about constant at \SI{2.5}{G}. The total magnetic flux encompassed increased to \SI{2.1e20}{Mx}, which is a result of the spatial growth of the dimming region encompassing a larger number of pre-existing photospheric magnetic elements. 

During this birth phase, the average intensity levels of the dimming in the AIA~\SI{193}{\angstrom} and \SI{211}{\angstrom} channels, respectively, decreased further to \SI{34}{\percent} and \SI{33}{\percent} of their initial quiet-Sun values. However, the average AIA~\SI{171}{\angstrom} intensity increased by about \SI{20}{\percent}. This increase occurs broadly throughout the coronal dimming. In addition, a small, bright coronal loop system appeared in the AIA~\SI{171}{\angstrom} channel precisely at the time of the filament eruption, as shown in Fig.~\ref{fig:overview}(f). This loop was apparent before the eruption in the hotter AIA 193~\AA\ channel, as indicated in Fig.~\ref{fig:overview}(e). A DEM analysis of the temporal evolution of the loop system showed that the average temperature decreased from a pre-eruption level of about~\SI{1.5}{MK} to a post-eruption level of about~\SI{1.1}{MK}, while the EM stayed roughly constant. Thus, the plasma in this loop system cooled significantly during the eruption, while its plasma content stayed approximately constant.

We checked whether the loop had an effect on our general coronal dimming DEM results and found only minor quantitative differences, with changes of approximately \SI{10}{\percent} in the derived EM and temperature, while no qualitative differences were observed. Therefore, the EM and temperature curves shown in Fig.~\ref{fig:timelines}(d) are also representative of the dimming plasma with this small loop system excluded. 

Aside from the influence of this loop system, the overall evolution of the coronal dimming plasma during the birth phase continued the trend from the precursor phase. During the birth phase, the EM of the dimming decreased from $1.4$~to~\SI{1.1e26}{cm^{-5}} and its average temperature from $1.5$~to~\SI{1.3}{MK}. This temperature decrease continued to follow the trend expected by adiabatic cooling, as shown in Fig.~\ref{fig:timelines}(d).

\subsection{Phase 2: Migration}

\begin{figure}
    \centering
    \includegraphics[width=\linewidth]{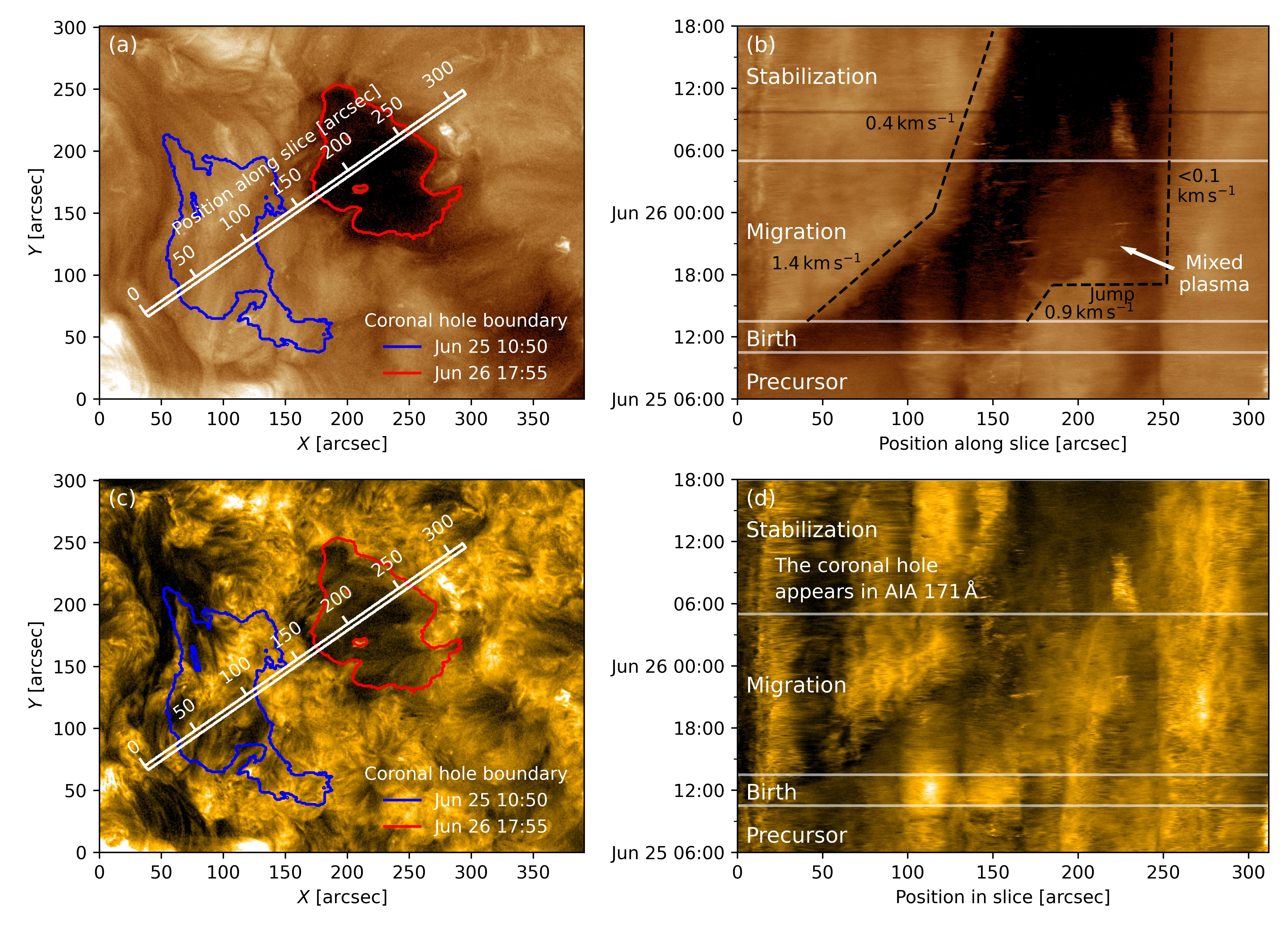}
    \caption{Migration of the coronal hole as seen in AIA~\SI{193}{\angstrom} (top row) and AIA~\SI{171}{\angstrom} (bottom row). The first column shows overview images from June~26~17:55. The coronal hole boundary from June~25~10:50, which we rigidly rotated with the solar surface to~17:55, is outlined in blue. This boundary corresponds to the initial location of the coronal dimming at the solar surface. The coronal hole boundary from June~26~17:55 is outlined in red. This boundary corresponds to the location where the coronal hole eventually stabilized. A slice in the migration direction is marked in white. The second column shows the temporal evolution of AIA intensities along the slice, describing the migration of the coronal hole.}
    \label{fig:stackplot}
\end{figure}

Starting from June~25~13:30, i.e., when the dimming has fully developed and it reached its largest size, we call the dimming a coronal hole. Over the following $15.5$~hours until June~26~05:00, this coronal hole migrated about \SI{150}{\arcsec} towards the north-west relative to the rotating solar surface. The migration phase started with an expansion of the coronal hole toward the south-west (13:00 -- 14:30), shown in Fig.~\ref{fig:overview}(g). Next, it shrank from the south (14:30 -- June 26~00:00) and an hour later from the east (15:30 -- June 26~00:00). Subsequently, it evolved and expanded to the north-west (15:00 -- June 26~05:00). On June~25 at 18:00, the coronal hole connected for the first time to what will become its final location, which is shown in Fig.~\ref{fig:overview}(h). On June 26~05:00, the coronal hole settled in its final location and its area stabilized.

In the movie associated with Fig.~\ref{fig:overview}, the coronal hole is clearly seen to drift across photospheric magnetic elements, which remained quasi-stationary on the solar surface. Because coronal holes host an open, quasi‑radial magnetic field, this drift motion must have involved continuous opening of closed field lines at the leading edge and simultaneous closing of open flux at the trailing edge. Moreover, since all coronal magnetic field lines must remain rooted in the photosphere, these opening and closing processes necessarily occurred via interchange reconnection at both edges to preserve the photospheric connectivity.
In addition, the coronal hole also drifted across the bright AIA~\SI{171}{\angstrom} loop system, which became visible in the AIA~\SI{171}{\angstrom} channel during the eruption (see Sect.~\ref{sec:formation}). This loop system did not drift with the coronal hole but remained fixed at the solar surface. Therefore, although initially within the coronal hole boundaries, this loop system was not part of the coronal hole.

In Figure \ref{fig:stackplot}, we analyze the drift speed quantitatively by evaluating the evolution of the AIA~$193$ and \SI{171}{\angstrom} intensities along a slice in the direction of the apparent motion. The slice co-rotates with the Sun. Distances are measured along the direction of the apparent motion so that small position values in the slice represent the plasma behind the coronal hole and large values the plasma at its front. At the beginning of the migration phase, the apparent drift speed of its leading boundary of \SI{0.9}{km\,s^{-1}} was smaller than the drift speed of its trailing boundary, which was \SI{1.4}{km\,s^{-1}}. This led to a contraction of the coronal hole in the north-west direction, which may have caused the observed expansion to the south. On June~25, at about~18:00, a sudden jump in the location of its leading boundary occurred, which corresponds to its first connection to its final location. This extension of the coronal hole to its final location appears somewhat brighter in the EUV, indicating that it is a mixture of coronal hole and quiet-Sun plasma. Finally, the leading boundary stayed fixed relative to the solar surface, while the trailing boundary further contracted to its final location.

Comparing the timelines in Figs.~\ref{fig:timelines} and~\ref{fig:timelines2}, we find that the derived coronal hole properties differ during the migration phase depending on whether the compact or wider coronal hole boundaries are used. The difference arises because the compact boundaries exclude the mixed plasma region in Fig.~\ref{fig:stackplot}(b), where the coronal hole first connects to its final location, whereas the wider boundaries include this region.
Focusing on the compact boundary timelines in Fig.~\ref{fig:timelines}, which capture only the coronal hole core, we observe that during the migration phase the EM decreased to \SI{0.8e26}{cm^{-5}} and the temperature dropped to \SI{1.1}{MK}. The temperature decrease follows adiabatic cooling. Meanwhile, the area and average magnetic flux density fluctuated without a clear physical pattern.
In contrast, the wider boundary timelines in Fig.~\ref{fig:timelines2} show that the temperature decrease deviates from adiabatic cooling. This is likely because the wider boundaries encompass not only clear coronal hole plasma but also mixed coronal hole and quiet-Sun plasma. However, the wider boundaries reveal a negative correlation between the coronal hole area and its average magnetic flux density. More precisely, as the coronal hole moved from a weaker to the stronger magnetic field region, its area decreased. This suggests that the shaded region may be part of the coronal hole but has not yet cooled to typical coronal hole temperatures. If this interpretation is correct, the magnetic flux evolution in the timelines of the wider boundaries implies that the total magnetic flux increased by at least \SI{20}{\percent} during the migration phase. This also means that the coronal hole did not solely migrate through interchange reconnection at its boundaries, which would conserve the total magnetic flux of the coronal hole, but that it gained additional magnetic flux during its migration.

\subsection{Phase 3: Stabilization}


Starting from June~26~05:00, the coronal hole settled at the location where the magnetic field model indicates open magnetic field lines. From then, its north-west boundary stayed fixed relative to the solar surface, while its south-east boundary continued to contract toward the center of the coronal hole. Simultaneously, the coronal hole slightly expanded along the perpendicular, north-east, south-west axis and finally stabilized on June~26~15:00. During this stabilization phase, the coronal hole became more pronounced in the AIA~\SI{193}{\angstrom} observations and also started to clearly appear in the AIA~\SI{171}{\angstrom} observations, shown in Fig.~\ref{fig:stackplot}.

The timelines in Figs.~\ref{fig:timelines} and~\ref{fig:timelines2} show that the evolution of the coronal hole properties appears to be smoother during the stabilization phase. Using the wider boundaries, the area, magnetic flux density, and magnetic flux of the coronal hole remained about constant, while when using the compact boundaries, all of these properties appeared to increase. This difference is due to the region of mixed coronal-hole and quiet-Sun plasma marked in Fig.~\ref{fig:stackplot}(b). As this region was already included in the wider boundaries during the migration phase, its darkening during the stabilization phase has no effect on the associated timelines. However, as this region was not part of the compact boundaries during the migration phase, its darkening causes it to become gradually included in the compact boundaries during the stabilization phase, resulting in the corresponding observed increase in area and magnetic flux. In both timelines, the EM now decreased exponentially by \SI{3}{\percent\,h^{-1}}, which might be related to a plasma outflow towards interplanetary space. The  measured temperature appeared to slightly increases from June~26~09:00 to~18:00 by about \SI{0.1}{MK}, which is probably an artifact of the DEM; the DEM curve at June~26~17:55 in Fig.~\ref{fig:timelines}(c) shows a weak peak at \SI{2.2}{MK}. This might  arise from residual calibration errors as such temperatures are physically unlikely in a coronal hole \citep{hofmeister2025}. 

Starting from June~26~18:00, when the coronal hole is at a longitude of \SI{20}{\degree} west of the central meridian, the coronal hole becomes harder to discern in the glare of the surrounding structures. This impairs us from analyzing its further evolution in a robust manner. However, the merged coronal hole can be discerned for at least another \SI{2.75}{days}, i.e., at least until \SI{4}{days} after its formation, while it rotated towards the solar limb, as shown in Fig.~\ref{fig:overview}(j). At the next solar rotation close to the solar disk center, the merged coronal hole seems again apparent, as shown in Fig.~\ref{fig:overview}(k)-(l). By comparing the latitudes of the merged coronal hole between the two solar rotations, we find that the portion corresponding to the newly formed coronal hole remains well recognizable. In contrast, the part corresponding to the pre-existing south-western coronal hole appears stretched toward the south-west and somewhat diminished, yet still present. The continued visibility of the coronal hole sets a lower bound on its lifetime of one solar rotation.

\section{Discussion} \label{sec:discussion}

Our study raises several interesting questions: 1)~When is a dark structure actually a coronal hole?  2)~Are there more coronal holes that are formed by filament eruptions? 3)~Why did the coronal hole migrate? 4)~Why is adiabatic cooling observable? We briefly discuss these questions here. 

\subsection{When is a dark structure actually a coronal hole?}
Clearly, not every dark structure on the Sun is a coronal hole. In the EUV, coronal holes can easily be confused with halos around active regions, coronal dimmings, and sometimes even filament channels. Active region halos can be distinguished spectroscopically from coronal holes \citep{lezzi2023}, while filament channels can be distinguished from coronal holes by the underlying photospheric magnetic field configuration \citep{delouille2018}. Coronal dimmings are usually distinguished from coronal holes by their association with CMEs, their shorter life times of \SI{8}-\SI{11}{hours} on average \citep{reinard2008, krista2017, krista2022}, and their dynamic evolution in coronal imagery, which features a fast recovery to quiet-Sun intensities \citep{Attrill:2008, Ronca_2024}. 
However, spectroscopically, many similarities between coronal holes and coronal dimmings have been reported. These include the reduced coronal densities and emission as compared to the ambient quiet-Sun values, plasma outflow along the approximately radial magnetic field, and non-thermal line broadening indicating Alf\'ven waves propagating towards interplanetary space \citep[review by][]{tian2021}.

Based on the findings of \citet{mcintosh2009}, our observations in this study, and the spectroscopic similarities between coronal dimmings and coronal holes described above, we argue that coronal dimmings might sometimes develop into coronal holes. The magnetic field strength in the footpoints of quiet-Sun filaments is similar to that in coronal holes, as both are rooted in photospheric magnetic elements \citep{delouille2018}. The magnetic field direction in filament legs becomes approximately radially aligned when the filament erupts and produces a CME, resulting in a local field orientation as in coronal holes. And the coronal density in the stretched filament leg might reach similarly low values as in coronal holes, as the CME evacuates plasma from the filament. In such a configuration, in the filament leg, wave–particle interactions may lead to energy dissipation and solar wind acceleration analogous to the processes in coronal holes. This would explain the observed plasma outflows in coronal dimmings. If these conditions then move to a location where the vertical magnetic field lines are preferably open, i.e., locations where magnetic field models indicate open magnetic field lines, the coronal dimming with the associated solar wind acceleration might stabilize and become persistent. Hence, the coronal dimming could transform into a real coronal hole. 

This scenario probably applies to the coronal hole under investigation. It was formed by a filament eruption with a subsequent coronal dimming. After its formation, it moved about \SI{150}{\arcsec} north-west within \SI{15.5}{hours} to a location where magnetic field models indicate open magnetic field lines, stabilizing the vertical magnetic fields. At this location, the EUV intensity did not then recover as would have been expected from coronal dimmings. Instead, the EUV intensity slowly decreased further, which fits to a persistent plasma outflow towards interplanetary space. With a lifetime of at least $1$~solar rotation, the structure is certainly not a coronal dimming anymore but became a small coronal hole.

\subsection{Are there more coronal holes that are formed by filament eruptions?} \label{subsec:morechs}
 \begin{figure}
    \centering
    \includegraphics[width=\linewidth]{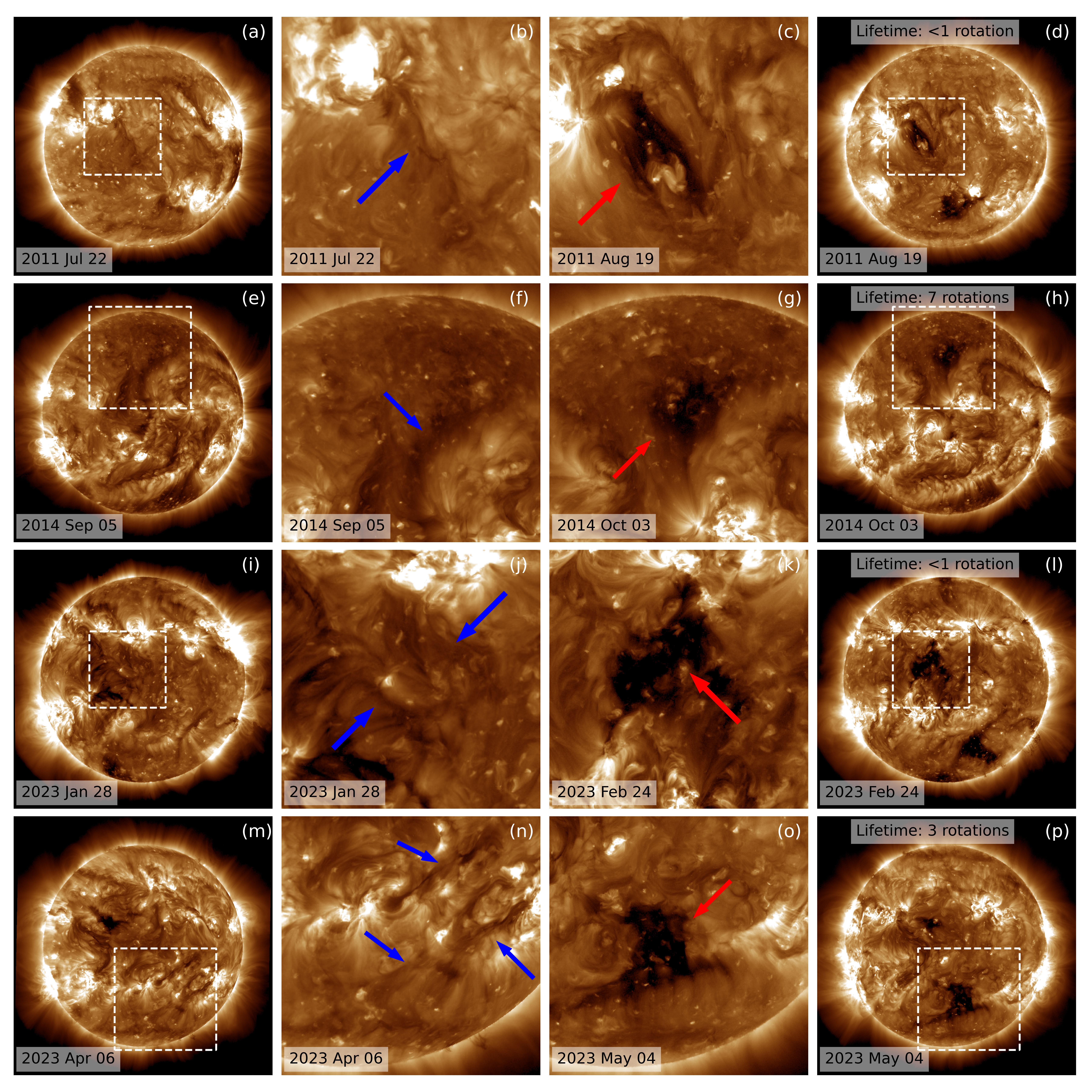}
    \caption{Additional coronal holes likely formed by filament eruptions. All panels display AIA ~\SI{193}{\angstrom} observations. First column: overview images showing filament channels, with the field of view for the second column marked in white. Second column: zoom-in on the filaments, indicated by blue arrows. Third column: same field of view at the next solar rotation. The filaments now disappeared and coronal holes (red arrows) have formed. Fourth column: overview images at the times of the coronal holes.}
    \label{fig:further_chs}
\end{figure}

\begin{table}[t]
    \centering
    \begin{tabular}{cc|cc|cc}
        Date & Position & Date & Position & Date & Position \\ \hline
          2011 Aug 19 & \SI{10}{\degree}N \SI{14}{\degree}E  & 2013 Dec 22 & \SI{38}{\degree}N \SI{8}{\degree}E  & 2023 Feb 22 & \SI{49}{\degree}S \SI{2}{\degree}W  \\ 
        2011 Aug 19 & \SI{31}{\degree}S \SI{5}{\degree}W  & 2014 Oct 03 & \SI{41}{\degree}N \SI{1}{\degree}W & 2023 May 04 & \SI{38}{\degree}S \SI{8}{\degree}W \\ 
        2012 Feb 15 & \SI{2}{\degree}S \SI{12}{\degree}E & 2016 Feb 08 & \SI{15}{\degree}S \SI{6}{\degree}E & 2023 Jun 08 &  \SI{30}{\degree}N \SI{4}{\degree}W\\ 
        2012 May 27 &\SI{46}{\degree}N  \SI{12}{\degree}E & 2016 Apr 27 & \SI{18}{\degree}N \SI{14}{\degree}E & 2023 Aug 12 & \SI{34}{\degree}N \SI{13}{\degree}E \\ 
        2012 Oct 24 &  \SI{29}{\degree}N \SI{7}{\degree}E& 2016 Jun 19 & \SI{19}{\degree}S \SI{8}{\degree}E & 2023 Nov 12 & \SI{50}{\degree}N  \SI{23}{\degree}W \\ 
        2012 Dec 25 & \SI{47}{\degree}S \SI{5}{\degree}E & 2016 Nov 12 &  \SI{11}{\degree}N \SI{12}{\degree}W& 2023 Dec 11 &  \SI{8}{\degree}S \SI{9}{\degree}E\\ 
        2013 Feb 11 &  \SI{32}{\degree}N \SI{8}{\degree}E& 2021 Sep 21 &  \SI{0}{\degree}N \SI{3}{\degree}E& ~ & ~ \\ 
        2013 May 25 & \SI{37}{\degree}N \SI{14}{\degree}W & 2022 Jul 26 &  \SI{13}{\degree}N \SI{14}{\degree}E& ~ & ~ \\ 
    \end{tabular}
    \caption{Date and location of coronal holes that might have formed from filament eruptions.}
    \label{tab:further_chs}
\end{table}

\citet{fainshtein2004} studied the locations of coronal holes and reported two populations: one in plage regions located away from photospheric magnetic inversion lines and another near these inversion lines. These two populations likely correspond to two different formation mechanisms: a continuous process driven by the gradual opening of magnetic field lines in plage regions and an eruptive process driven by filament eruptions. However, direct observations of these formation processes are rare. To our knowledge, the only clear observation of a coronal hole forming through the slow, continuous process is a coronal hole on 2012~September~3 reported by \citet{inglis2019}. In addition to the coronal hole studied here, a formation of a coronal hole by a filament eruption has been reported by \citet{heinemann2018a}, which lasted for $10$~solar rotations.

Building on these studies, we searched the AIA database at a cadence of one week from 2010~to~2024 for coronal holes likely formed by filament eruptions. We stepped through the AIA~\SI{193}{\angstrom} images and looked for images where a clear filament channel was visible in one image and where the filament channel disappeared and a clear coronal hole nearby is observed instead one solar rotation later. In total, we identified $22$~of these events, which we list in Table~\ref{tab:further_chs}. Figure~\ref{fig:further_chs}, presents four of the clearest cases of this list. In each row, the first column shows the location of a large filament channel on the Sun, and the second column provides a zoomed-in view. The third and fourth columns show the same location during the next solar rotation, where the previously visible filament channel has been replaced by a medium-sized coronal hole. The disappearance of the filament channel and the appearance of the coronal hole nearby strongly suggest that these coronal holes were formed by filament eruptions. The lifetimes of these coronal holes range from $<1$~to~$7$~solar rotations. 

The number of coronal holes we identified as forming from filament eruptions likely represents a lower limit, as both filament channels and coronal holes, particularly at mid-to-high latitudes, are often partially obscured by surrounding quiet-Sun loops. This observational limitation complicates the identification of both structures and, consequently, the association of coronal hole formations with filament eruptions. Considering this selection effect, we would not be surprised if the number of coronal holes that have formed by filament eruptions is substantially larger than the number of cases we have identified. 

\newpage
\subsection{Why did the coronal hole migrate?}

The migration of the coronal hole over \SI{150}{\arcsec} into a region with PFSS-modeled open magnetic field lines shows that coronal holes do not always originate in areas of modeled open magnetic fields. We find that they can form in areas where models predict closed magnetic field lines, such as a coronal dimming close to filaments, open the magnetic field at that location, and then migrate to regions with modeled open magnetic field lines. The migration to a region with modeled open magnetic field lines stabilizes the coronal hole and can be interpreted as a minimization of energy to keep the magnetic field open. To observe the open field lines in our study, we had to set the source surface height of the PFSS model, i.e., the altitude at which the magnetic field lines are considered open, to a low value of $1.6$~solar radii. If we had used higher source surface heights, such as the standard values of $2$~or~$2.5$~solar radii, the model would not have shown open field lines in the region where the coronal hole settled. Theoretically, the source surface height should be set to large heights so as to not artificially constrain the modeling domain. However, in practice, source surface heights as low as this, or even lower, are often needed to at least roughly match the EUV and solar wind observations \citep{lee2011, asvestari2019, badman2020, panasenco2020}. These low source surface heights are likely needed to compensate for systematic errors in the model, which include inaccuracies in full-disk magnetograms \citep{plowman2020}, incomplete magnetic field evolution in synoptic magnetograms \citep{yang2024, heinemann2025}, oversimplified PFSS assumptions that assume that the corona is static, potential, and force-free, as well as artifacts introduced by the selected numerical schemes used to compute the PFSS solution. In addition, the need for such a low source surface height to obtain a modeled open magnetic field region at the location of the observed coronal hole might indicate that this open magnetic field region is not very stable.

During the migration phase to the region of modeled open magnetic field, the coronal hole drifted across the solar surface, above regions where the underlying photospheric magnetic elements were quasi-stationary. Analogous observations have been reported by \citet{heinemann2023}, who studied the dissolution of a coronal hole due to its drift into a large quiet-Sun loop arcade. Both our study and theirs demonstrate that there is no stable connection between the manifestation of a coronal hole in the corona and the underlying photospheric elements. Instead, the coronal magnetic field in the coronal holes appears to have reconnected with the photospheric magnetic field as needed, primarily following the open coronal magnetic field topology. 

This tendency to follow the large‐scale open‐field structure is also reflected in another well‐known property of coronal holes, i.e., their typically more-rigid rotation rates as compared to typical quiet‐Sun structures. \citet{sheeley1987} proposed, based on analytical calculations and numerical simulations, that supergranular diffusion acts as a meridional flux‐transport mechanism, counteracting the shear from differential rotation and producing a more rigid rotation of the large‐scale coronal magnetic field. As manifestations of this field, coronal holes follow the large‐scale open‐field topology, which hence governs their rotation as well as their migration into regions with a preferred open‐field configuration, as demonstrated in this study.

\subsection{Why is adiabatic cooling observable?}
During the formation and migration phases, the evolution of the $T$–EM relationship followed well the behavior expected from adiabatic cooling. However, adiabatic cooling assumes no net heat exchange with the surrounding, implying that either all other heating and cooling processes are negligible or that they cancel out.

The assumption that all other heating and cooling processes are negligible is difficult to reconcile with the known energy balance of the corona. The characteristic timescale for radiative cooling is tens of minutes \citep{aschwanden2005, landi2012}, meaning that without heating, the plasma would cool rapidly from \SI{1.6} to \SI{0.6}{MK} in tens of minutes. To maintain a roughly constant temperature, heating must supply energy at a matching rate. Assuming a characteristic heating timescale of \SI{20}{minutes}, this implies that during the \SI{1.25}{day} interval in our study, in which we seem to observe adiabatic cooling, the corona experiences approximately 90 heating episodes, each of which provides enough energy to raise the coronal temperature by about \SI{1}{MK}. Heating processes are therefore substantial and cannot be neglected.

The alternative is that non-adiabatic heating and cooling processes are precisely balanced throughout the entire period during which the temperature evolution seems to follow adiabatic cooling. However, this would imply that coronal heating and non-adiabatic cooling mechanisms merely compensate for each other and thus do not drive the temperature of the corona, which seems an implausible scenario. Therefore, it remains unclear why the observed $T$–EM evolution follows so closely the trend expected from adiabatic cooling.



\section{Summary} \label{sec:summary}

We have studied the formation of a coronal hole from a large quiet-Sun filament eruption near the solar disk center from 2014~June~25 to June~26 using SDO observations. The formation progressed in the following steps:
\begin{itemize}
    \item A filament eruption from a quiet-Sun filament caused a coronal dimming.
    \item This dimming developed into a coronal hole, which moved about \SI{150}{\arcsec} to the north-west over \SI{15.5}{hours}, crossing a small coronal loop system and drifting above the solar surface in which the photospheric magnetic elements remained quasi-stationary.    
    \item     Throughout this evolution, the $T$-EM relationship of the forming coronal hole followed the trend expected from adiabatic cooling.
    \item The coronal hole then settled at a location where a PFSS model indicated open magnetic field lines. During this stabilization, the AIA intensities and EM began to decrease further.
    \item After about \SI{1.5}{days}, starting from a longitude of approximately \SI{20}{\degree}~west, the coronal hole began to be obscured by the brighter surrounding quiet-Sun loops, but remained identifiable. About this time, it merged with a pre-existing nearby coronal hole at its south-west.
    \item At the next solar rotation, the newly formed  portion of the coronal hole was still visible at the same location. This sets its lifetime to $>1$~solar rotation.
\end{itemize}

The detailed evolution observed in this event shows that coronal holes can, under certain conditions, form as a consequence of quiet-Sun filament eruptions. The structure persisted for at least one solar rotation, migrated to a region of modeled open magnetic field, and showed a progressive decrease in EM, which is consistent with the expectations of coronal hole formation. The drift of the coronal hole across the photosphere, along with the absence of a stable connection to specific magnetic elements, points to a continuous magnetic restructuring during the migration phase that is likely governed by interchange reconnection.

Motivated by the event reported here, we searched the AIA database for additional cases in which a quiet-Sun filament eruption may have led to the formation of a coronal hole. We identified $22$~events between 2010 and 2024, of which we presented $4$ of the clearest cases that produced large coronal holes with lifetimes from~$<1$ and up to $7$~solar rotations. Owing to observational limitations and selection effects, the number of events identified likely represents a lower bound. Given that filament eruptions are frequently observed to modify existing coronal hole boundaries, causing them grow, shrink, or shift rapidly, it would not be surprising if a larger number of coronal holes are actually formed by filament eruptions.%

~\\
\textit{Acknowledgements:} Authors past the first author are listed alphabetically, not by contribution. SJH, AK, MH, and DWS acknowledge support by the National Science Foundation (NSF) grants AGS-2229100 and AST-2005887. SGH acknowledges funding from the Finnish Research Council project SWATCH (343581) and the Austrian Science Fund (FWF) Erwin-Schrödinger fellowship J-4560. EA acknowledges support from the Finnish Research Council (Research Fellow grant number 355659). VJ acknowledges support to the NASA Postdoctoral Program at the Goddard Space Flight Center, administered by Oak Ridge Associated Universities under contract with NASA.


\facility{SDO(AIA), SDO(HMI)}
\software{Astropy \citep{astropy1,astropy2,astropy3},  
    Matplotlib \citep{matplotlib},
    Numpy \citep{numpy}, 
    Scipy \citep{scipy}, 
    Sunpy \citep{sunpy},
    SolarSoftWare (SSW) \citep{freeland1998},
    DEM inversion \citep{cheung2016}, 
    PSF tools \citep{hofmeister2024a, hofmeister2025}}

\bibliographystyle{aa} 
\bibliography{bibliography}

\end{document}